# First-principles modeling of the polycyclic aromatic hydrocarbons reduction

D. W. Boukhvalov [1,*] X. Feng[2] and K. Müllen[2]

[1]*School of Computational Sciences, Korea Institute for Advanced Study (KIAS) Hoegiro 87, Dongdaemun-Gu, Seoul, 130-722, Korean Republic*

[2]*Max Planck Institute for Polymer Research, Ackermannweg 10, 55128 Mainz, Germany*

*Density functional theory modelling of the reduction of realistic nanographene molecules ($C_{42}H_{18}$, $C_{48}H_{18}$ and $C_{60}H_{24}$) by molecular hydrogen evidences for the presence of limits in the hydrogenation process. These limits caused the contentions between three-fold symmetry of polycyclic aromatic hydrocarbon molecules and two-fold symmetry of adsorbed hydrogen pairs. Increase of the binding energy between nanographenes during reduction is also discussed as possible cause of the experimentally observed limited hydrogenation of studied nanographenes.*

E-mail: danil@kias.re.kr

## 1. Introduction

Hydrogenation of aromatic hydrocarbons and related carbon materials is an actual topic in modern experimental (see Ref. [1] and references therein) and computational[2,3] chemistry. Fabrication of the hydrogenated graphene monolayer- namely graphane[4] has stimulated the interests for the functionalization of the nanographenes. Further experimental results[5] suggest that for the significant sample dependency of the hydrogenation processes of graphene.[4] Previous calculations[2,6] discuss the leading role of the edges for the hydrogenation process of graphene nanoribbons. The similarity of atomic structure between graphene[7] and polycyclic aromatic hydrocarbons (PAH)[8,9] suggests that defined PAH could be a suitable model for the modeling of graphene hydrogenation. Theoretical prediction of the intriguing electronic and magnetic properties of the graphene and graphene nanoribbons,[10] nanoenginering of graphene by partial hydrogenation[11], manufacturing of several graphenic nanoribbons from PAHs[9] and experimentally detected limits in PAH reduction by catalytic hydrogenation[1] also require detailed survey of the hydrogenation of PAH with given shape.

The previous works[12-14] suggests that density functional theory (DFT) is the powerful tool for the modeling of realistic chemistry of compounds with a honeycomb lattices[4]. In the present work we studied step by step reduction of hexa-peri-hexabenzocoronene (HBC, $C_{42}H_{18}$ **1,** the structure is shown in Fig. 1) and two other graphene molecules with different shapes ($C_{48}H_{18}$ **2** and $C_{60}H_{24}$ **3**, shown in Fig. 2 and 3 respectively). For check the possibility of stack formation between partially hydrogenated PAH molecules in process of hydrogenation calculation of the binding energies for the initial, intermediate and final studies of functionalization had been performed.

## 2. Computational method and model

The modeling was carried out by density functional theory (DFT) realized in the pseudopotential code SIESTA,[15] as was done in our previous works.[4,12,16] For the hydrogenation processes all calculations were done using the generalized gradient approximation (GGA-PBE)[17] which is the most suitable for the description of graphene-adatom chemical bonds.[4,6,12] Full optimization of all atomic positions was performed. During the optimization, the electronic ground state was found self-consistently using norm-conserving pseudo-potentials for cores and a double-$\zeta$ plus polarization basis of localized orbitals for carbon and oxygen, and double-$\zeta$ basis for hydrogen. Optimization of the forces and total energies was performed with an accuracy of 0.04 eV/Å and 1 meV, respectively. All calculations were carried out for an energy mesh cut off of 360 Ry and a k-point mesh 8×8×1 in the Mokhorst-Park scheme.[17] All calculations have been performed in spin polarized mode. Use of this mode for the knowingly non-magnetic configurations is the important issue of the modeling of graphene and related systems functionalization. Incorrect initial structures should provide magnetic configurations after several loop of the self-consistent process. Calculations in this mode could help the computational time usually spent due to very probable misprints in the starting configurations.

For the realistic modeling of the processes of catalytic hydrogenation by molecular hydrogen on each step of reduction process the pair of hydrogen atoms was added from same side on nearest carbon atoms (*ortho* position) connected with double bond. This structure is corresponding to the boat like structure (Fig. 4a). Previous calculations[17] suggest that the most energetically favorable is chair like structure (Fig. 4b). In our modeling we take into account possible reconstruction from boat-like to chair-like structure at the final steps of hydrogenation. The graphenes molecules have been placed in the center of the empty cubic box with the side length of 4 nm. Formation energy was calculated by standard formula $E_{form} = E_{step\ N+1} - (E_{step\ N} - E_{H2})$, where $E_{step\ N}$ and $E_{step\ N+1}$ is the total energies of the studied graphene molecules before and after chemisorption of the pair of hydrogen atoms, $E_{H2}$ is the total energy of molecular hydrogen in empty box. Calculation of the formation energy with using total energy of molecular hydrogen is necessary for the simulation of the realistic hydrogenation processes.

The calculations of the atomic structure and binding energies for the physisorption of benzene molecules as simplest of molecules with carbon hexagon with π-π bonds on PAH with different level of hydrogenation are done using the LDA functional[18] which is reliable for the description of graphene adsorbed molecule interaction without chemical bond formations.[16,21] The value of binding energies between graphene molecules and benzene molecules calculated with using standard formula $E_{bind} = E_{g+bz} - (E_g + E_{bz})$, where $E_{g+bz}$ is the total energy of graphene molecules with adsorbed benzene molecule, $E_g$ and $E_{bz}$ - the total energies of graphene and benzene molecule in the same box.

## 3. Results and discussions

We have started the modeling of HBC from calculation of the adsorption of hydrogen pairs from one site on different sites (see Fig. 5a, b). Similar to previous theoretical results[2,6,12] for relevant graphitic compounds the formation energy for hydrogenation of nanographene edges is negative and for chemisorption of the hydrogen pair in the central part is positive (about 0.3 eV, near to the similar value for bulk graphene[12]). For the modeling of the next steps of reduction we have calculated the formation energies of the different probable positions of hydrogen pairs (see for example of the first step of hydrogenation of **1** Fig. 5c) and used the configuration with minimal total energy as starting point for the next step. It is necessary to note that for all steps of hydrogenation of all three studied compound total energy of the most favorable configuration is at least 0.1 eV lower than the other. These results evidence for the rather high probability of the reaction pathway reported on Fig. 1.

There are two issues that seem as controversial with the experimental results[1] for the reduction of **1**. In experiments the process of hydrogenation requires additional light heating but all steps of the reduction reported in theory are exothermic. It is necessary to note that our model is quite ideal and exactly valid only for the gaseous phase.[22] In contrast with realistic processes all sites on **1** could be reached by hydrogen molecules. In experiments the heating has been used for the transport of hydrogen molecules to the graphene surface. Therefore, the results of our calculations could be used as examination of the limits of reduction when exothermic processes turn to be endothermic and could be used as estimation of the temperatures required for the further steps of reduction.

Theoretical results predict significant level of hydrogenation of **1** but in the experiment reported the reduction only of the edge peripheries (see Fig. 1). The cause of the difference between experimental and theoretical results could be due to the changes in the binding energies between molecules of **1**. For modeling of these changes we performed the calculation of the binding energies between studied graphene molecules at different levels of the hydrogenation and benzene molecule (minimal molecule with hexagon structure and double carbon-carbon bonds, see Fig. 4). The results of calculations (see Tab. 1) prove significant enhancement of the binding between all types of graphene molecules and benzene after hydrogenation of the edges. Further reduction leads to the increase of the binding energies. Nature of these changes of the energy of adsorption of benzene molecule on graphenes is in distortions of graphene sheet caused by hydrogenation.[6,12] The corrugations of graphene (see for example Fig. 4a, b) sheet significantly increase the binding energies between graphene and toluene[16] and should provide formation of stack-like structures.[23] The "conglutination" of the graphene molecules makes the central part of the molecules unachievable for the hydrogen in contrast to the planar graphene under hydrogen plasma treatment[4,5] and could be the limit of the further reduction of graphene molecules.

Similar to **1** we examined the energetics of the reduction of graphene molecules **2** and **3**. For **2** the total reduction after exothermic step of the hydrogen adhesion are reported (see Fig. 2), for **3** similarly to **1** the significant level of the reduction could be obtained. For understanding the nature of the limit in hydrogenation due to turn exothermic to endothermic process taking into account the symmetry of the graphene is required. In contrast to the square like graphenic compounds or graphene nanoribbons and nanographenes[6] all studied molecules has six- or three-fold symmetry but in process of reduction we add on each step only pair of the hydrogen atoms. These pairs lead local two-fold symmetry appearance. The contention between significant distortions of nanographene flat (see Fig. 6), local symmetry near chemisorbed pair and symmetry of the molecule results changes of the chemisorption energies of each step of reductions (see Figs. 1-3). Proposed explanation is work for **1** and **3** but for the totally reducible **2** require taking account of additional source of chemical activity. In contrast to **1** and **3** the graphene molecule **2** has not only armchair but also ziz-zag edges. Larger chemical activity of this type of the graphene edges have been discussed in the several theoretical works.[6,20] The presence of these types of edges in **2** enhance chemical activity of this type of graphene and leads vanishing of discussed above limit in the hydrogenation.

For check the possible role of the reconstruction of the boat-like structure obtained in our calculations (Figs. 4a, 6) to the most energetically favorable for graphene chair-like structure (Fig. 4b) we performed calculation of the total energy for the case of the reconstruction of the obtained final configurations and find that the energy gain is about 70 meV/H for all three structures. This result is near to reported for the flat infinite graphene.[17] For check the influence of this reconstruction of hydrogen adatoms we have performed the calculations for the probable next steps of reduction and fond that similarly to the boat-like structure for the chair-like structure this processes also should be endothermic.

## 4. Conclusions

Performed DFT modeling of the reduction of polycyclic aromatic hydrocarbons explain the difference between hydrogenation of nanographenes and free standing quasi-infinite graphene sheets. For the realistic nanographenes shape of the samples play crucial role. The contentions between three-fold symmetry of PAH molecules and two-fold symmetry of adsorbed hydrogen pair is the case of appearance of additional energy barriers for the total hydrogenation. Increasing of the binding energy between partially reduced nanographenes makes the central areas of these molecules unreachable for the hydrogen molecules. Taking into account these two main limitations in nanographenes functionalization is necessary for the modeling of graphene nanoribbons covalent functionalization.

**References**

1. Watson, M. D.; Debije, M. G.; Warman, J. M.; Müllen, K. *J. Am. Chem. Soc.* **2004**, *126*, 766.

2. Lin, Y.; Ding, F.; Yakobson, B. I. *Phys. Rev. B* **2008**, *78*, 041402; Sheka, E. F.; Chernozatonskii, L. A. *Int. J. Quant. Chem.* **2010**, *110*, 1938; Sheka, E. F.; Chernozatonskii, L. A. *JETP* **2010**, *110*, 121.

3. Shahab Naghavi, S.; Gruhn, T.; Alijani, V.; Fecher, G. H.; Felser, C.; Medjanik, K.; Kutnyahov, D.; nepijko, S. A.; Schönhense, G.; Rieger, R.; Baumgerten, M.; Müllen, K. *J. Mol. Spectr.* in press.

4. Elias, D. C.; Nair, R. R.; Mohiuddin, T. G. M.; Morozov, S. V.; Blake, P.; Halsall, M. P.; Ferrari, A. C.; Boukhvalov, D. W.; Katsnelson, M. I.; Geim, A. K.; Novoselov, K. S. *Science*, **2009**, *323*, 610.

5. Luo, Z.; Yu, T.; Kim, K.-J.; Ni, Z.; You, Y.; Lim, S.; Shen, Z.; Wang, S.; Lin, J. *ACS Nano* **2009**, *3*, 1781.

6. Boukhvalov, D. W.; Katsnelson, M. I. *Nano Lett.* **2008**, *8*, 4378; Kosimov, D. P.; Dzhurakhalov, A. A.; Peeters, F. M. *Phys. Rev. B* **2010**, *81*, 195414; Ao, Z. M.; Hernández-Nieves, A. D.; Peeters, F. M.; Li, S. *Appl. Phys. Lett.* **2010**, *97*, 233109.

7. Casiraghi, C.; Hartschuh, A.; Quian, H.; Piscanec, S.; Georgi, C.; Fasoli, A.; Novoselov. K. S.; Basko, D. M.; Ferrari, A. C. *Nano Lett.* **2009**, *9*, 1433; Neubeck, S.; You, Y. M.; Ni, Z. H.; Blake, P.; Shen, Z. X.; Geim, A. K.; Novoselov. K. S. *Appl. Phys. Lett.* **2010**, *97*, 053110; Xu, Y. N.; Zhan, D.; Liu, L.; Suo, H.; Nguyen, T. T.; Zhao, C.; Shen, Z. X. *ACS Nano* in press.

8. Wu, J.; Pisula, W.; Müllen, K. *Chem. Rev.* **2007**, *107*, 718; Riger, R.; Müllen, K. *J. Phys. Org. Chem.* **2010**, *23*, 315.

9. Cai, J.; Ruffieux, P.; Jaafar, R.; Bieri, M.; Braun, T.; Blankenburg, S.; Muoth, M.; Seitsonen, A. P.; Saleh, M.; Feng, X.; Müllen, K.; Fasel. R. *Nature* **2010**, *466*, 470; Yang, X. Y.; Dou, X.; Rouhanipour, A.; Zhi, L. J.; Räder, H. J.; Müllen, K. *J. Am. Chem. Soc.* **2008**, *130*, 4216.

10. Singh, A. K.; Yakobson, B. I. *Nano. Lett.* **2009**, *9*, 1540; Xiang, H.; Kan, E.; Wei, S.-H.; Whangbo, M.-H.; Yang, J. *Nano Lett.*, **2009**, *9*, 4025.

11. Chernozatonskii, L. A.; Sorokin, P. B. *J. Phys. Chem. C* **2010**, *114*, 3225; Chernozatonskii, L. A.; Sorokin, P. B.; Bruning, J. W. *Appl. Phys. Lett.* **2007**, *91*, 183103; Chernozatonskii, L. A.; Sorokin, P. B.; Belova, E. E.; Bruning, J.; Fedorov, A. S. *JETP Lett.* **2007**, *85*, 77.

12. Boukhvalov, D. W.; Katsnelson, M. I.; Lichtenstein, A. I. *Phys, Rev. B* **2008**, *77*, 035427.

13. J. M. Soler, E. Artacho, J. D. Gale, A. Garsia, J. Junquera, P., Orejon, and D. Sanchez-Portal, *J. Phys.: Condens. Matter* **2002**, *14*, 2745.

14. Boukhvalov, D. W *Surf. Sci.* **2010**, *604*, 2190.

15. Perdew, J. P.; Burke, K.; Ernzerhof, M. *Phys. Rev. Lett.* **1996**, *77*, 3865.

16. Monkhorst, H. J.; Park, J. D. *Phys. Rev. B* **1976**, *13*, 5188-92.

17. Sofo, J. O.; Chaudhari, A. S.; Barber, G. D. *Phys. Rev. B* **2007**, *75*, 153401.

18. Perdew, J. P.; Zunger, A. *Phys. Rev. B* **1981**, *23*, 5048.

19. Ataca, C.; Aktürk, E.; Ciraci, S. *Phys. Rev. B* **2009**, *79*, 041406; Ataca, C.; Aktürk, E.; Ciraci, S.; Ustunel, H. *Appl. Phys. Lett.* **2008**, *93*, 043123; Ao, Z. M.; Peeters, F. M. *Phys. Rev. B* **2010**, *81*, 205406; Leenaerts, O.; Partoens, B.; Peeters, F. M. *Phys. Rev. B* **2009**, *79*, 2235440; Leenaerts, O.; Partoens, B.; Peeters, F. M. *Microel. J.* **2009**, *40*, 860; Leenaerts, O.; Partoens, B.; Peeters, F. M. *Appl. Phys. Lett.* **2008**, *92*, 243125; Leenaerts, O.; Partoens, B.; Peeters, F. M. *Phys. Rev. B* **2008**, *77*, 125416; Wehling, T. O.; Novoselov, K. S.; Morozov, S. V.; Vdovin, E. E.; Katsnelson, M. I.; Geim, A. K.; Lichtenstein, A. I. *Nano Lett.* **2008**, *8*, 173; Wehling, T. O.; Katsnelson, M. I.; Lichtenstein, A. I. *Appl.Phys. Lett.* **2009**, *93*, 202110; Ribeiro, R. M.; Peres, N. M. R.; Coutinho, J.; Briddon, P. R. *Phys. Rev. B* **2008**, *78*, 075442; Zanella, I.; Guerini, S.; Fagan, S. B.; Mendes Filho, J.; Souza Filho, A. G. *Phys. Rev. B* **2008**, *77*, 073404; Cordero, N. A.; Alonso, J. A. *Nanotechnology* **2007**, *18*, 485705; Widenkvist, E.; Boukhvalov, D. W.; Rubino, S.; Akhtar, S.; Lu, J.; Quinlan, R. A.; Katsnelson, M. I.; Leefer, K.; Grennberg, H.; Jansson, U. *J. Phys. D: Appl. Phys.* **2009**, *42*, 112003.

20. Rouhanipour, A.; Roy, M.; Feng, X.; Räder, H. J.; Müllen, K. *Angew. Chem. Int. Ed.* **2009**, *48*, 4602.

21. Yang, S. B.; Feng, X.; Wang, L.; Tang, K.; Maier, J.; Müllen, K. *Angew. Chem. Int. Ed.* **2010**, *49*, 4795; Räder, H. J.; Rouhanipour, A.; Talarico, A. M.; Palermo, V.; Samorì, P.; Müllen, K. *Nat. Mater.* **2006**, *5*, 276; Laursen, B. W.; Norgaard, K.; Reitzel, N.; Simonsen, J. B.; Nielsen, C. B.; Als-Nielsen, J.; Bjornholm, T.; Solling, T. I.; Nielsen, M. M.; Bunk, O.; Kjaer, K.; Tchebotareva, N.; Watson, M. D.; Müllen, K.; Piris, J. *Langmuir* **2004**, *20*, 4139.

**Table I** Binding energies in meV/$C_6H_6$ for adsorption of benzene molecules from one a two sides (in parenthesis) of pure, intermediate, and final level of the reduction of **1**, **2** and **3** compounds (see Figs. 1-3).

| Compound | Pure | Intermediate reduction | Final reduction |
|---|---|---|---|
| **1** | 67 (87) | 119 (153) | 236 (312) |
| **2** | 123 (122) | 187 (191) | 429 (497) |
| **3** | 117 (123) | 161 (197) | 351 (379) |

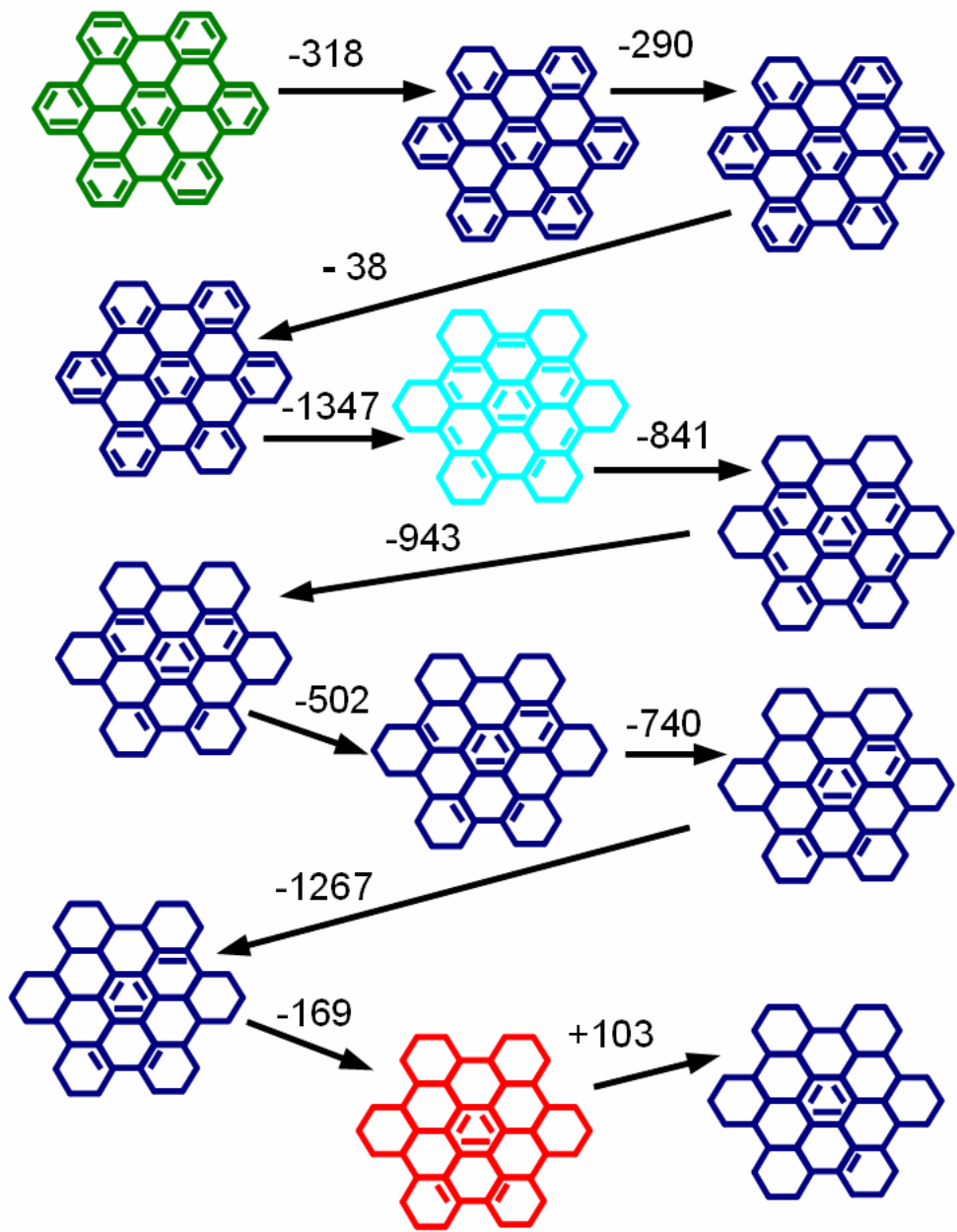

**Figure 1** A sketch of step by step reduction of hexabenzocoronene ($C_{42}H_{18}$, **1**). All energies are shown in meV/$H_2$. Initial structure is indicated by green, intermediate stable structure obtained in the experiment[1] (see description in text) by cyan and final by red.

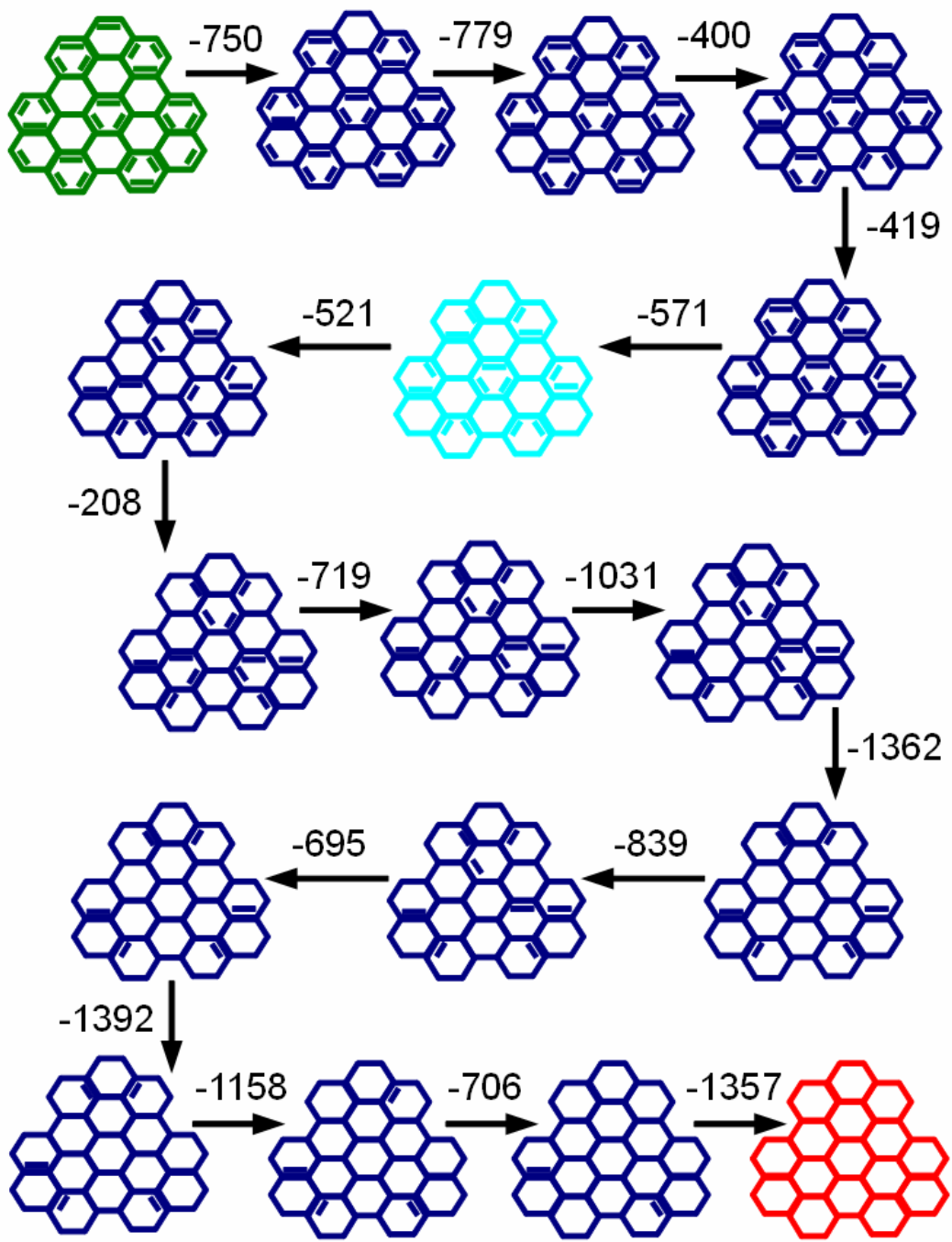

**Figure 2** A sketch of step by step reduction of $C_{48}H_{18}$ (**2**). All energies are shown in meV/$H_2$. Initial structure is indicated by green, intermediate stable structure (see description in text) by cyan and final by red.

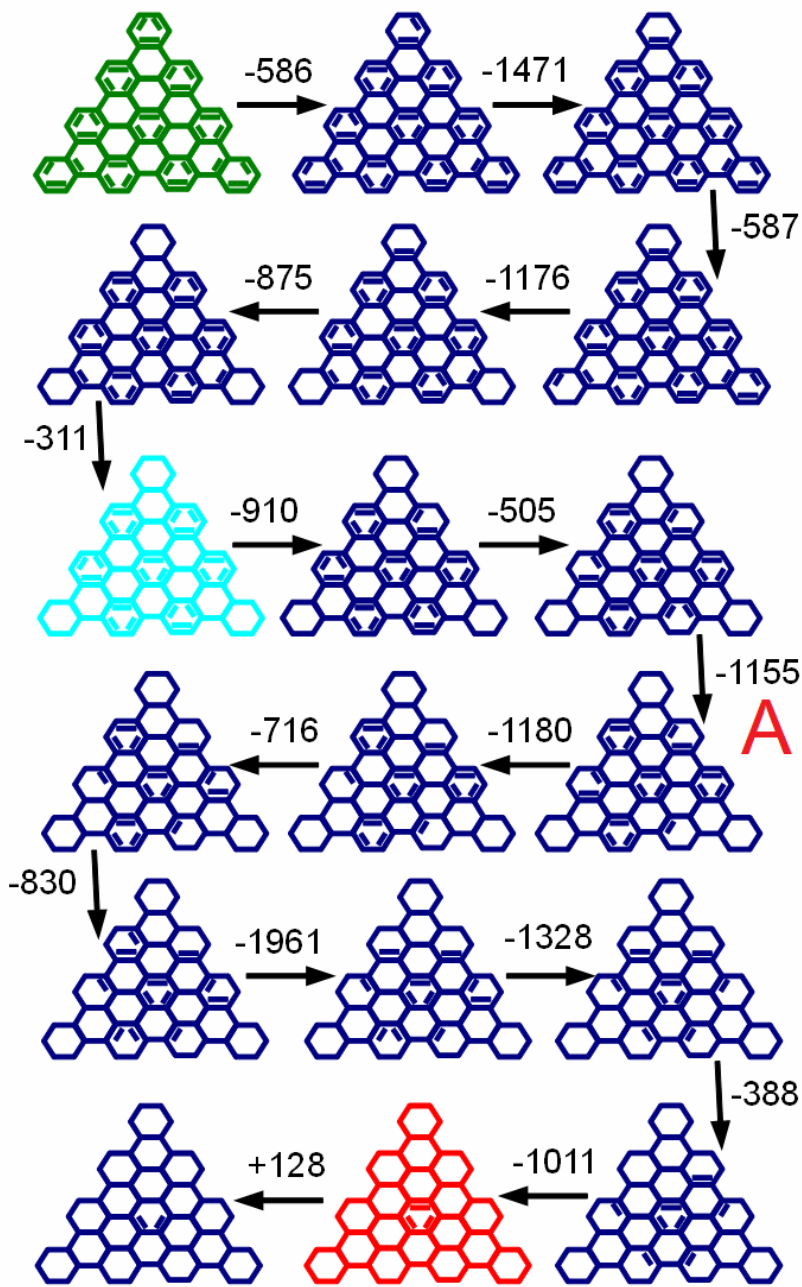

**Figure 3** A sketch of step by step reduction of $C_{60}H_{24}$ (**3**). All energies are shown in meV/$H_2$. Initial structure is indicated by green, intermediate stable structure (see description in text) by cyan and final by red. Optimized atomic structure of the configuration pointed by the letter A is shown on Fig. 6.

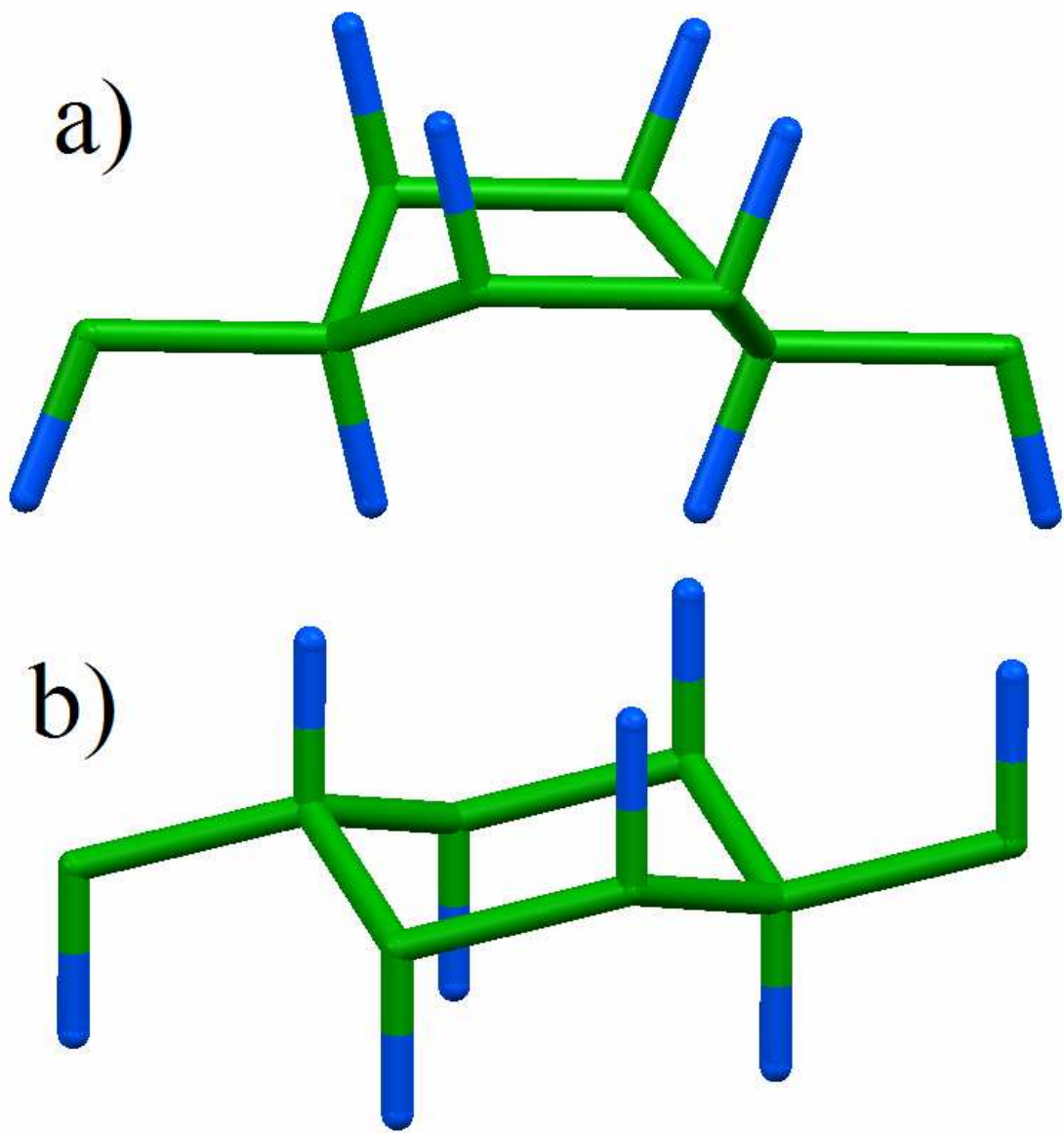

**Figure 4** Optimized atomic structures of "boat" (a) and "chair" (b) atomic structures of 100% hydrogenated graphene.

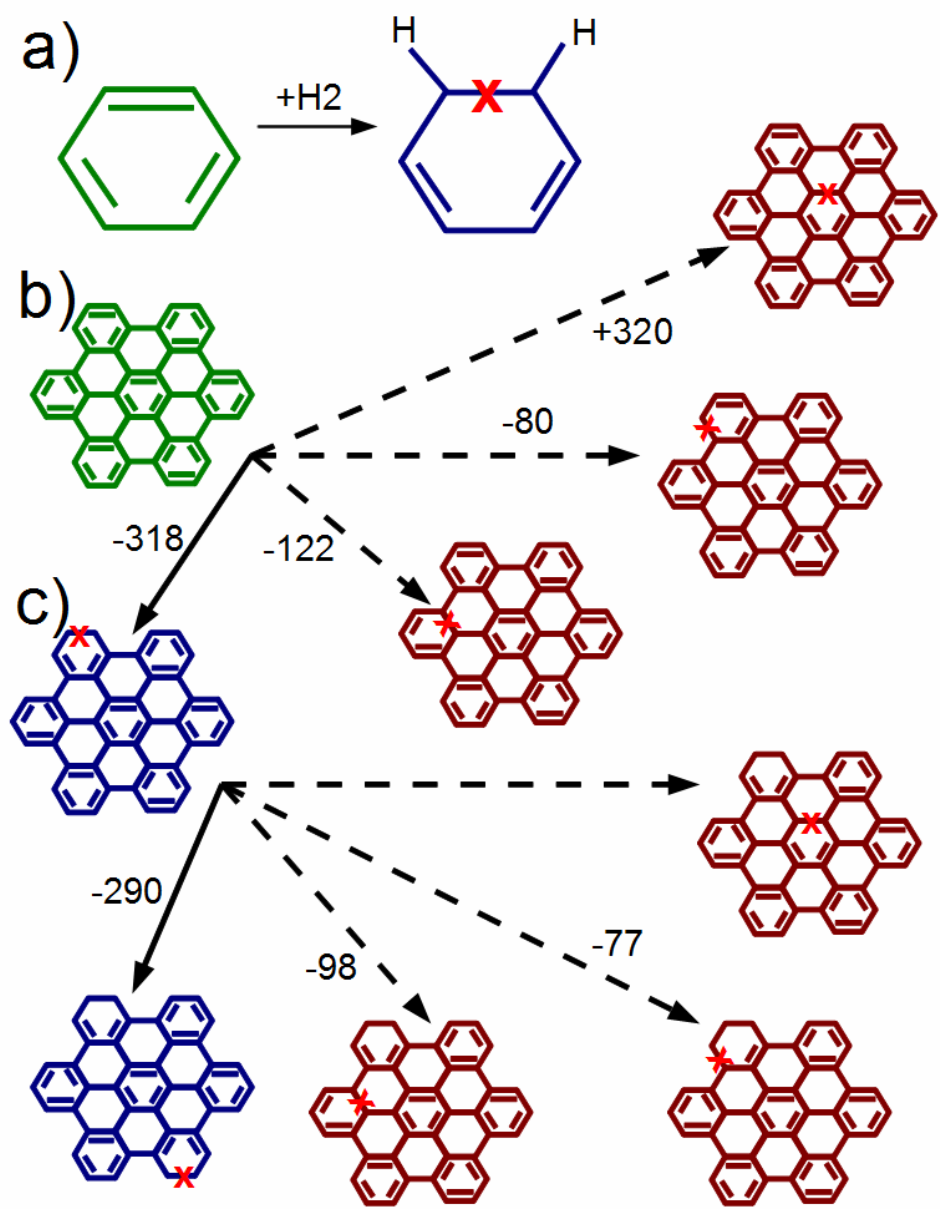

**Figure 5** A sketch of the modeling of the hydrogen pair chemisorption on the carbon hexagon (a) with change of double C=C bond to single (shown by the red cross); and modeling process of two first steps of reduction of hexabenzocoronene ($C_{42}H_{18}$, **1**). All energies are shown in meV/$H_2$. Initial configurations are shown by green, most probable configurations by blue and less energetically favorable by brown.

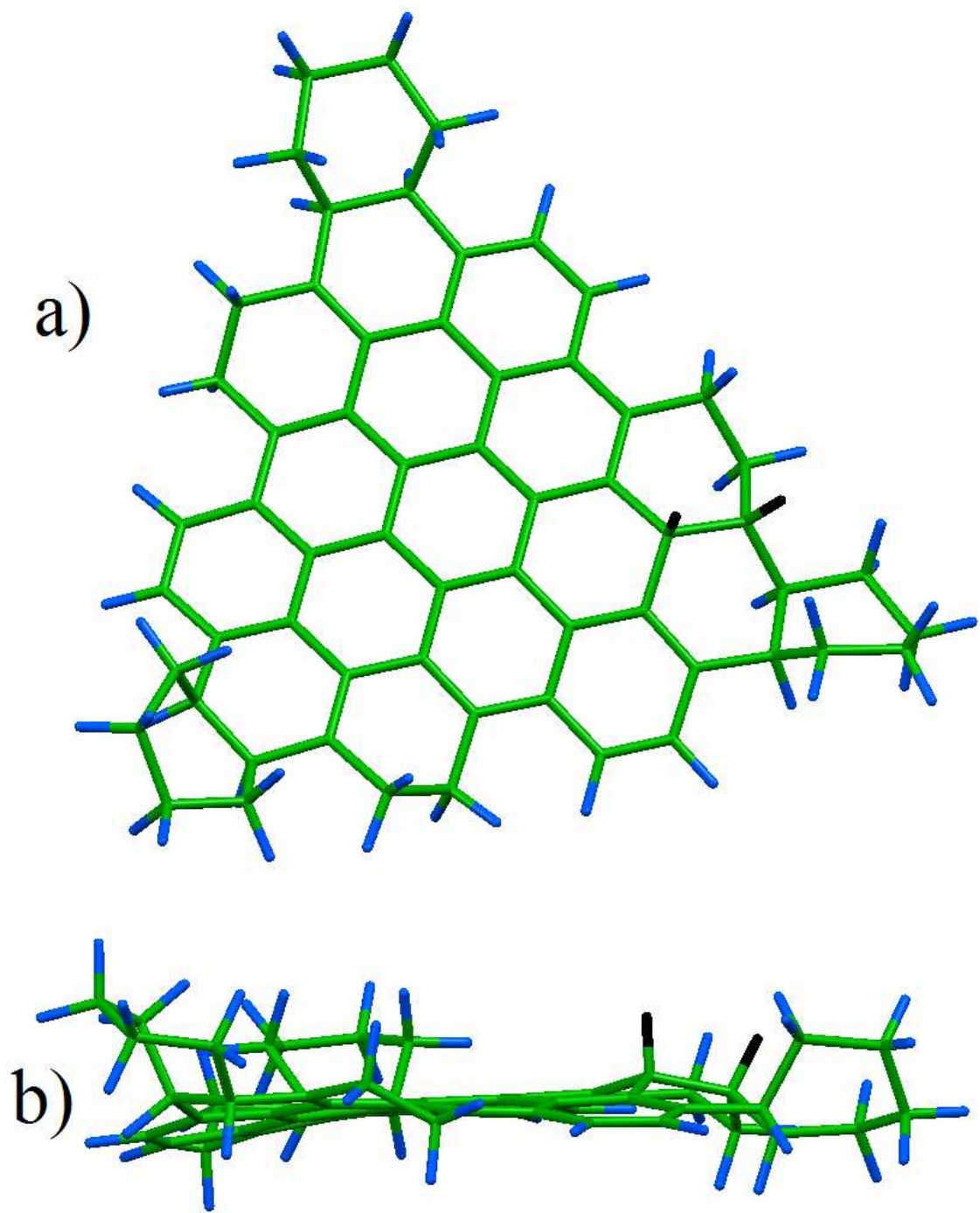

**Figure 6** Optimized atomic structure of the intermediate step of the hydrogenation of PAH molecule indicted by letter A on Fig. 3.

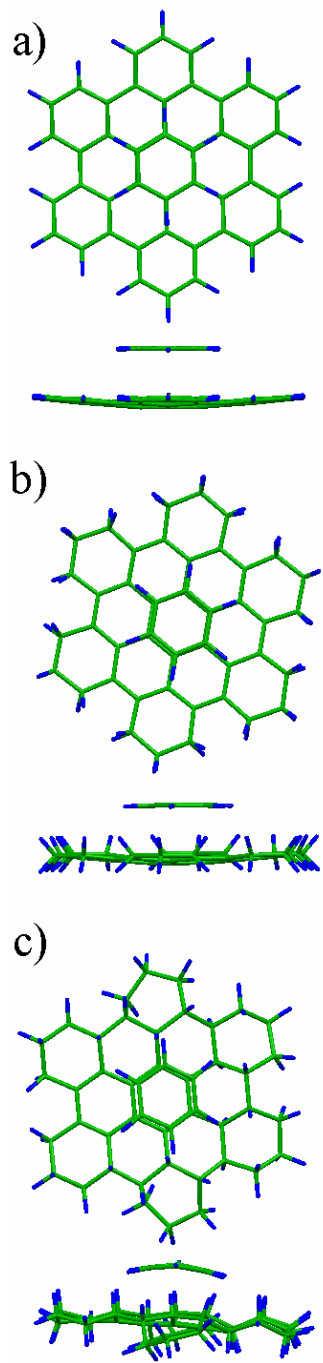

**Figure 7** Optimized atomic structures of benzene molecule adsorbed on (a) pure hexabenzocoronene $C_{42}H_{18}$; (b) hexa-peri-hexabenzocoronene $C_{42}H_{36}$, and (c) hexabenzocoronene with maximal level of hydrogenation ($C_{42}H_{50}$).